# THE EFFECT OF TEMPERATURE ON BASAL METABOLISM OF *MNEMIOPSIS LEIDYI*

JAMILEH JAVIDPOUR[1], EDUARDO RAMIREZ-ROMERO[2] AND THOMAS LARSEN[3]

[1] Department of Biology, University of Southern Denmark, Campusvej 55, 5230 Odense M, Denmark; jamileh@biology.sdu.dk

[2] Fish Ecology Group, Instituto Mediterráneo de Estudios Avanzados, IMEDEA (CSIC-UIB), C/Miquel Marqués 21, 07190 Esporles, Illes Balears, Spain; edu.ramirezromero@gmail.com

[3] Max Planck Institute for the Science of Human History, Kahlaische Str. 10, 07745 Jena, Germany; larsen@shh.mpg.de

ABSTRACT. To evaluate the influence of temperature on metabolic performance on the invasive ctenophore *Mnemiopsis leidyi*, we exposed fully acclimatized adults to conditions typical for the annual variability of the Western Baltic Sea region. We derived basal metabolic rates from oxygen consumption rates of adult *M. leidyi* specimens exposed to temperatures between 3.5º to 20.5°C at a salinity of 22. We found a Q10 value of 3.67, which means that the carbon specific respiration rates are about 9 times greater at 20ºC than 3ºC. According to this rate, a small-sized individual 20 mm in oral-aboral length, would without feeding have enough nutrient reserves to survive 80 days at 3°C, but only 9 days at 20ºC. Thus, prey availability during late summer is critical for *M. leidyi* population survival.

## 1. INTRODUCTION

We assessed the basal metabolic rates of *M. leidyi* under temperature conditions typical for the Western Baltic Sea region (Figure 1). The basal metabolic rate is the energy required by an organism to maintain vital functions, without growing, digesting, reproducing or any activities that demand additional energy. Oxygen consumption rate is considered as a good parameter to approximate the metabolic rate of a heterotrophic organism because they mainly gain their energy by oxidizing carbon compounds to carbon dioxide and water. For poikilothermic metazoans, temperature is the main abiotic factors influencing basal metabolic rates, which means that starvation conditions are more critical for survival of *M. leidyi* during summer than winter months [1]. To obtain quantitative data for *M. leidyi* carbon specific respiration rates, we exposed adult individuals to temperatures between 3.5º and 20.5°C, a temperature range that is 7°C colder than previously reported [2].

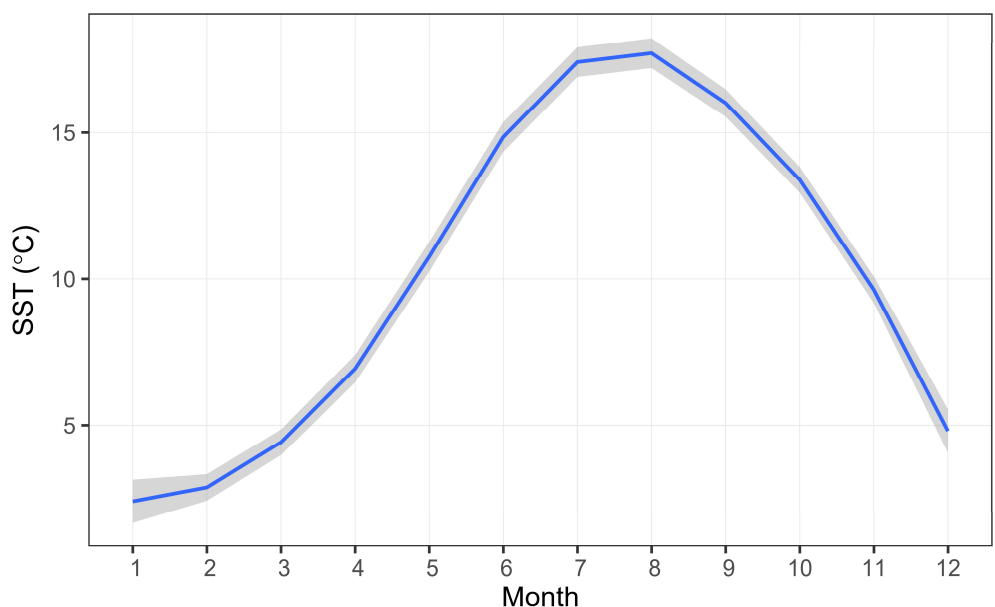

**Figure 1**. Mean sea surface temperatures from 2000 to 2010 in Kiel Bight. The grey areas represent the 95% confidence interval across the 11 years. Note that the temperatures for overwintering *M. leidyi* may differ from SST since the animals seek seasonal refuge in deeper waters. SST data were obtained from the AVHRR Pathfinder Version 5.2 (PFV5.2), US National Oceanographic Data Center and GHRSST (http://pathfinder.nodc.noaa.gov). The PFV5.2 data are an updated version of the Pathfinder Version 5.0 and 5.1 collection described in Casey et al. [3]. Monthly averaged images with 4 km spatial resolution were used and SST series was extracted for the Kiel Bight.

To obtain experimental animals, we collected wild specimens in January 2007 when the ambient water temperature was 3°C and the salinity 22, and subsequently acclimatized them to the different temperatures. To infer the relationship between total body length and wet mass, we used a size range of 0.5-30 mm. The *M. leidyi* population of the western Baltic is dominated by small sized individuals of 20±0.3 mm according to Javidpour, Molinero, Peschutter and Sommer [4]. We derived basal metabolic rates from oxygen consumption rates and bulk carbon and nitrogen contents relative to dry/wet mass and total length of *M. leidyi* specimens. From these measurements, we derived the carbon requirements and carbon turnover rates of *M. leidyi*. Since *M. leidyi* is an osmo-conformer over a wide range of salinities, we report the metabolic rate relative to carbon rather than to dry/wet mass to factor out non-respiring matter, i.e. salt.

## 2. MATERIAL & METHODS

**2.1. Collection of *Mnemiopsis leidyi*.** Adult ctenophores were collected twice from Kiel Fjord (54°19,7' N, 10°09,5' E) by vertically towing a WP2 net from 10 m depth (500 μm mesh-size with 0.8 m wide net) in January 2007. Healthy individuals were isolated immediately in a beaker at ambient temperature (3°C) and salinity (22). We tested the effect of temperature change on the carbon specific respiration rate ($Q_{10}$) at five temperature levels between 3.5° and 20.5°C.

The respiration experiment was carried out after a 24 h of acclimatizing the ctenophores to experimental conditions (active swimming and searching for food). Only undamaged adult specimens of about the same size (17.4± 1.1mm; n = 72) were selected for the experiment. Oxygen consumption to measure respiration rate of *M. leidyi*, two individuals where transferred into a 270 ml Winkler bottle, filled with 0.25 μm filtered and autoclaved seawater of incubation salinity and temperature. To guarantee oxygen saturation at the beginning of the experiment, seawater was aerated for ten minutes prior to the experiment. The bottles were sealed air-free and placed into climatic chambers for 24 hours. The temperature within the bottles was held constant within a range of +/- 0.25°C. The whole incubation period was performed in darkness. Each temperature treatment consisted of four replicates plus four controls (i.e. bottles without *M. leidyi*). Oxygen consumption was measured using the Winkler technique [5].

During the incubation period, ctenophores were checked for vitality by observation of the regular movement of the comb-rows. All individuals were alive at the end of incubation. *M. leidyi* was filtered with a 1 mm mesh and wet mass (WM) was determined. Specimens were then left to dry for 48 hours at 60°C to measure dry mass (DM). The lobes of *M. leidyi* were included for biomass estimation. Carbon and nitrogen contents were determined via combustion in an element analyser (Thermo Scientific Flash 2000).

**2.2. Data processing.** Oxygen consumption data were normalized to the milligrams of body carbon of the ctenophores and to the incubation time in order to obtain carbon specific respiration rate [6]. To convert oxygen consumption of zooplankton into carbon units, a respiratory quotient of 0.97 [7] was applied to transform the carbon specific respiration rate into carbon specific carbon consumption rate (μg of carbon metabolized per mg of carbon body content). For the temperature effect only, all results carried out at a salinity of 22 were combined to obtain a five-point exponential regression.

Statistical analysis was carried out with the software Statistica version 12.0. Factorial analysis of variance (ANOVA) was used to examine the effect of temperature on the respiration rate.

# 3. RESULTS

**3.1. Biometric Data.** The relationship between total length and WM is depicted in Figure 2. It can be expressed as follows:

WM = 0.0029*L$^{1.962}$ (eqn. 1),

where L is the total length including lobes in mm.

Oxygen consumption data were normalized to milligrams of body C and to the incubation time to obtain carbon specific respiration rate, i.e. µmol $O_2$ per mg C ind$^{-1}$ h$^{-1}$.

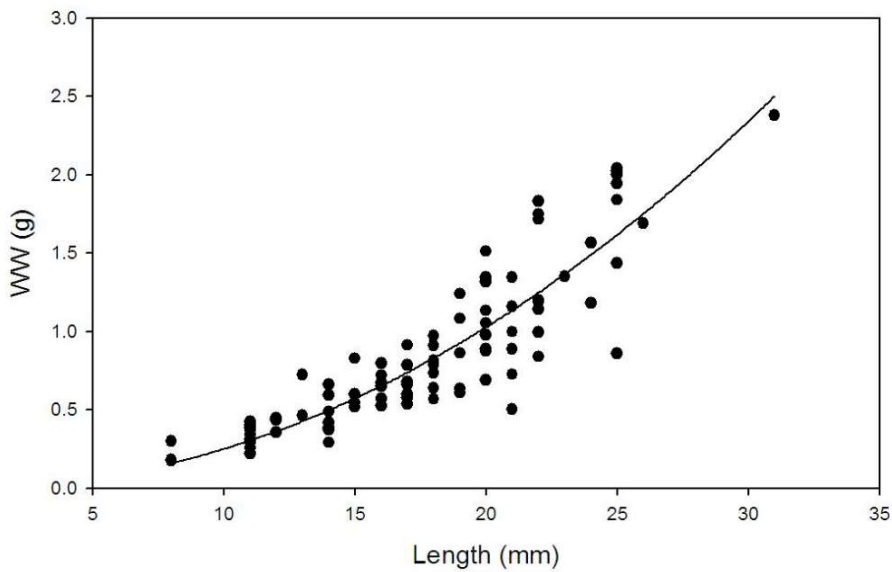

**Figure 2**. The relationship ($r^2_{adj}$= 0.855, n=84; p<0.01) between wet mass (WW; in grams) and total length (in millimetres) of *M. leidyi* in water with a salinity of 22.

**3.2. Q10.** For the temperature range of 3.5° to 20.5°C (Figure 2), we derived the following exponential regression:

Respiration (µmol $O^2$ / mg C ind$^{-1}$ h$^{-1}$) = 0.0298 $e^{0.13T°}$ (eqn. 2)

From eq. 2, we calculated a Q10 value of 3.67. This means that the respiration rate of an adult *M. leidyi* is approximately 9 times greater at 20º than at 3ºC.

**3.3. Energetic requirements.** At a salinity of 22, the carbon specific consumption rate can be expressed as follows:

$0.0298^{e0.13*T} * 0.97*12 = 0.347e^{0.13*T}$ (µg C per mg body C ind$^{-1}$ h$^{-1}$) (eqn. 3)

From this exponential relationship between carbon consumption and temperature, an adult *M. leidyi* turns over 4.6 µg C per mg body C ind$^{-1}$ h$^{-1}$ at 20ºC, and 0.51 µg C per mg body C ind$^{-1}$ h$^{-1}$ at 3ºC (Figure 3).

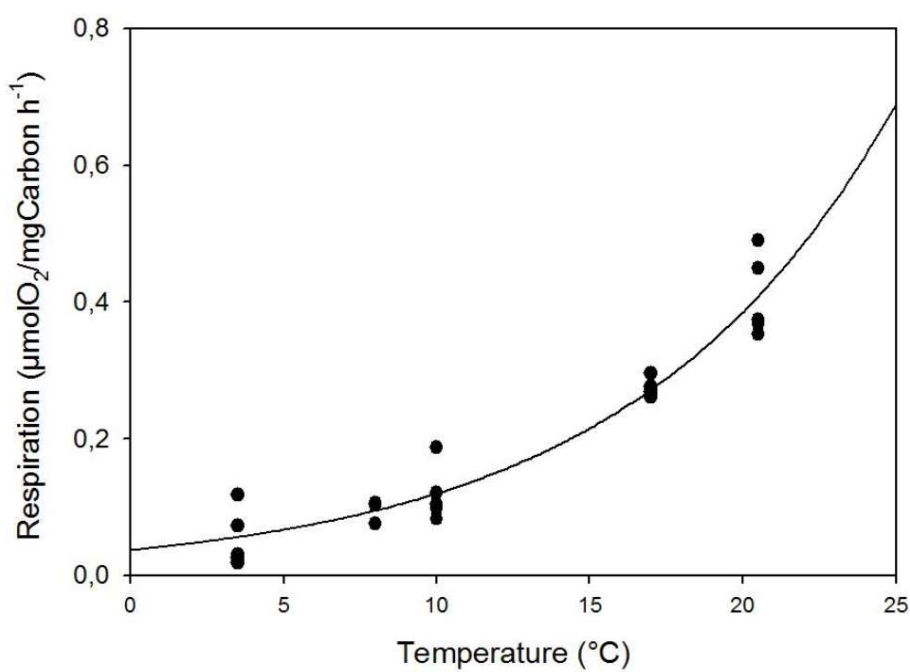

**Figure 3**. Exponential fit ($r^2_{adj}$=0.8; p<0.01) of the respiration rate in µmol O$^2$ per mg C h$^{-1}$ for the five temperature treatments with a salinity of 22.

## 4. Discussion

To our knowledge, we quantified for the first time the basal metabolic rates of *M. leidyi* at temperatures below 10ºC. Our results show that the rate follows the same exponential relationship for the 3.5° to 20.5°C range as for the 10° to 28ºC range previously reported for native populations [1, 8-10]. According to our results, typically sized adults have enough nutrient reserves to survive for 80 days at 3°C without feeding. It cannot be excluded that *M. leidyi* feed on benthic food sources during winter, but their feeding rates are likely to be much reduced compared to summer since adults shrink to larval size during winter in their native habitats [11]. However, our results show that non-feeding periods are much more critical for survival during the post-bloom phase in Kiel Bay when the water temperature can reach 20°C. At this temperature, adults have sufficient nutrient reserves to survive 9 days without feeding.